\documentclass[aps,twocolumn,superscriptaddress,nobalancelastpage,floatfix,showpacs]{revtex4}

\usepackage{graphicx}
\usepackage{latexsym}
\usepackage{longtable}
\usepackage{amsfonts}
\usepackage{amssymb}
\usepackage{dcolumn}

\begin{document}

\title {$^{57}$Fe M\"ossbauer study of magnetic ordering in superconducting
$\rm K_{0.85}Fe_{1.83}Se_{2.09}$ single crystals}

\author{D.H. Ryan}
\author{W.N. Rowan-Weetaluktuk}
\affiliation{Physics Department and Centre for the Physics of Materials,
McGill University, Montreal, H3A 2T8, Canada}

\author{J.M. Cadogan}
\affiliation{Department of Physics and Astronomy, 
University of Manitoba, Winnipeg, Manitoba, R3T 2N2, Canada}

\author{R. Hu}
\author{W.E. Straszheim}
\author{S.L. Bud'ko}
\author{P.C. Canfield}
\affiliation{Ames Laboratory, U.S. DOE and Department of Physics and Astronomy, 
Iowa State University, Ames, IA 50011, USA}

\date{\today}

\begin{abstract}
The magnetic ordering of superconducting single crystals of
$\rm K_{0.85}Fe_{1.83}Se_{2.09}$ has been studied between 10~K and
550~K using $^{57}$Fe M\"ossbauer spectroscopy. Despite being superconducting
below T$_{sc} \sim 30$~K, the iron sublattice in
$\rm K_{0.85}Fe_{1.83}Se_{2.09}$ clearly exhibits magnetic order
from well below T$_{sc}$ to its N\'eel temperature of
T$_N= \, 532 \, \pm \, 2$~K. The iron moments are ordered perpendicular
to the single crystal plates, {\it i.e.} parallel to the crystal
c-axis. The order collapses rapidly above 500~K and the accompanying
growth of a paramagnetic component suggests that the magnetic transition
may be first order, which may explain the unusual temperature
dependence reported in recent neutron diffraction studies.
\end{abstract}

\maketitle

\section{Introduction}
\label{intro}

The co-existence of magnetism and superconductivity in the
new iron-chalcogenide superconductors raises the possibility
of unconventional pairing mechanisms that may be associated
with their magnetism. \cite{Kenji, Lumsden, Canfield, Johnpierre}
As with the more established cuprate superconductors, the iron-based
superconductors have layered structures; the planar Fe layers tetrahedrally
coordinated by As or chalcogen anions (Se or Te) are believed to be
responsible for superconductivity. Stacking of the FeAs building blocks with
alkali, alkaline earth or rare earth-oxygen spacer layers forms the basic
classes of iron arsenic superconductors in these compounds: 111-type AFeAs
\cite{Wang}, 122-type $\rm AFe_2As_2$\cite{Rotter, Sefat, Ni, Jasper},
1111-type ROFeAs \cite{Kamihara, Chen} and more complex block
containing phases, e.g. $\rm Sr_2VO_3FeAs$ \cite{Zhu},
$\rm Sr_3Sc_2Fe_2As_2O_5$ \cite{Zhu2}, $\rm Sr_4Sc_2Fe_2As_2O_6$
\cite{Chen2}. The simple binary 11-type iron chalcogenides have no spacer
layers and superconductivity can be induced by doping FeTe with
S\cite{Rongwei} or Se\cite{Mizu}. Unlike the other iron-based
superconductors, FeSe is a superconductor \cite{Hsu}, T$_{sc}\sim 8$~K,\ with
no static magnetic order and its transition temperature can be increased up
to 37~K by applying pressure \cite{Med} or 15~K in $\rm FeSe_{0.5}Te_{0.5}$
\cite{Mizu}. More recently, superconductivity above 30 K has been reported in
$\rm A_xFe_{2-y}Se_2$ (A = K, Cs, Rb or Tl) \cite{Guo, Ying, Li, Fang,Hu1931},
a compound with the same unit cell structure as the $\rm AFe_2As_2$ compounds.

Recent neutron diffraction measurements on $\rm K_{0.8}Fe_{1.6}Se_{2}$
have suggested that not only does magnetic order co-exist with
superconductivity but that the iron moments are remarkably large
(3.31~$\mu_B$/Fe) and their magnetic ordering temperature may
be as high as 560~K \cite{Weibao0830}. The relatively complex antiferromagnetic
structure places all of the iron moments parallel to the c-axis.
Muon spin relaxation ($\mu SR$) has confirmed the high magnetic ordering
temperature in $\rm Cs_{0.8}(FeSe_{0.98})_2$ \cite{Sher} but the development
of a paramagnetic component near T$_N$ suggests that the transition may
be first order in nature rather than being a more conventional second
order transition. First order magnetic transitions are commonly associated
with changes in crystal structure, and both synchrotron x-ray diffraction
\cite{Pomj1919} and neutron diffraction \cite{Weibao0830, Weibao3674} have
now shown evidence for a structural change from $I4/m$ to $I4/mmm$
associated with a disordering of iron vacancies that occurs in the
vicinity of the magnetic
transition. However, the magnetic structure adopted by the iron
sublattice has not been confirmed, and room temperature neutron
diffraction studies of $\rm Cs_yFe_{2-x}Se_2$ \cite{Pomj1919} and
$\rm A_yFe_{2-x}Se_2$ (A = Rb, K) \cite{Pomj3380} have suggested that the iron
moments may be much smaller ($\sim$2.5~$\mu_B$/Fe) and also that the
magnetic structure may be far more complex than initially suggested, with
the iron atoms being distributed among two (magnetically) inequivalent
sublattices and carrying very different magnetic moments. Moreover,
even the ordering {\em direction} has been questioned and it is possible
that the iron moments may lie in the ab-plane, at least for
$\rm Cs_yFe_{2-x}Se_2$ \cite{Pomj1919}, rather
than parallel to the c-axis as initially suggested \cite{Weibao0830}.

Given the many questions surrounding the magnetic ordering of the
iron moments in the $\rm A_yFe_{2-x}Se_2$ system, we have undertaken
a $^{57}$Fe M\"ossbauer study $\rm K_{0.85}Fe_{1.83}Se_{2.09}$.
While M\"ossbauer spectroscopy cannot
be used to determine magnetic structures directly, it is a quantitative
local probe that can be used to set hard limits on possible structures.
As we will show below, the observation of a single, well-split magnetic
component allows us to rule out any structure in which the iron
sub-lattice is further subdivided into multiple, inequivalent sites,
and the scale of the splitting is inconsistent with  smaller
moment estimates. Furthermore, by working with single crystal
samples, we are able to demonstrate that the moments order parallel
to the c-axis, ruling out any models that invoke planar ordering.
Finally, the development of a paramagnetic component that co-exists
with the magnetically ordered phase and that grows at its expense
on heating through T$_N$, confirms that the magnetic transition is
indeed first order in nature.

\section{Experimental methods}
\label{exp}

The preparation and characterisation of the crystals used in
this study has been described in detail elsewhere \cite{Hu1931} so
only a basic description will be provided here.
Single crystals of $\rm K_{0.85}Fe_{1.83}Se_{2.09}$ were grown from
a $\rm K_{0.8}Fe_{2}Se_{2}$ melt, as described in Ref.\cite{Ying}.
First the FeSe precursor was
prepared by reacting stoichiometric Fe and Se at 1050$^{\circ}$C. Then, K and
FeSe with a nominal composition of $\rm K_{0.8}Fe_2Se_2$ were placed in
an alumina crucible that was sealed in an amorphous silica tube.
The growth was placed in a furnace in a
vented enclosure and heated to 1050$^{\circ}$C, where it was held for
a 2 hour soak. The furnace temperature was then slowly lowered
to 750$^{\circ}$C over 50 hours; the furnace was then turned off
and the sample ``furnace cooled'' over an additional 10 hours.
Once the ampoules were opened, large ($\sim 1\times 1\times 0.02$ cm$^3$)
dark shiny crystals could be mechanically separated from the solidified melt.

Crystals were characterised by powder x-ray diffraction using a Rigaku
Miniflex X-ray diffractometer.
The average chemical composition was determined by examining multiple points
on a cleaved surface of the crystal, using wavelength dispersive x-ray
spectroscopy (WDS) in a JEOL JXA-8200 electron microscope. A backscattered
electron analysis (BSE) was performed using an accelerating voltage of 20~kV.
Magnetic susceptibility was measured in a Quantum Design MPMS, SQUID magnetometer.
The x-ray diffraction pattern can be indexed using the I4/mmm space group, with
lattice parameters refined by Rietica of $a=3.8897(8)$\AA\ and
$c=14.141(3)$\AA, in good agreement with previous values \cite{Ying}.
The in-plane resistivity of the furnace cooled sample is
very similar to that of earlier reports \cite{Ying, Wangdm}:
a broad resistive maximum centered near 160~K is followed by a
drop of nearly a factor of 6 ($\rho_{300K}/\rho_{35K}$).
The sharp transition to zero resistance gives
an onset temperature of T$_{sc}=30.1$~K. Corresponding features
were also seen in the susceptibility and heat capacity.

The as-grown crystals were too thick to serve as M\"ossbauer
absorber but we found that they could be cleaved quite
easily using a razor blade, much like mica, and many large-area
plates could be formed from each crystal.
Two single crystal mosaic samples were prepared from the same batch
of crystals. The first, for low-temperature work, was prepared by
attaching several single crystal plates to a 12~mm diameter disc
of 100$\mu m$ thick Kapton foil using Apiezon N grease.
Care was taken to ensure that there were no gaps, but minimal
overlap between the crystals. This sample was transferred
promptly to a vibration-isolated closed-cycle refrigerator
with the sample held in vacuum. The second sample, for the
high-temperature work, was attached to a $\frac{1}{2}$-inch
diameter 10-mil beryllium disc using diluted GE-7031 varnish
before being mounted in a resistively heated oven, again with the
sample in vacuum. While we operated somewhat above the maximum
service temperature of the varnish, the sample was cycled above
250$^{\circ}$C three times without any evidence of degradation.

The M\"ossbauer spectra were collected on conventional
spectrometers using  50~mCi $^{57}$Co{\bf Rh} sources mounted
on electromechanical drives operated in constant acceleration
mode (on the high-temperature system) and sine-mode (on the
low-temperature system). The spectrometers were calibrated
against $\alpha -$Fe metal at room temperature. The closed-cycle
refrigerator cools to 10~K, with temperature sensing and control
using a calibrated silicon diode mounted on the copper sample
stage. Measured gradients (centre to edge of sample)
in the oven are less than 1~K up to 750~K. Control and sensing
rely on four type-K thermocouples. Temperature stability in both
cases is better than 0.2~K. Spectra were fitted using a conventional
non-linear least-squares minimisation routine to a sum of
equal-width Lorentzian lines. Magnetic patterns were fitted
assuming first-order perturbation in order to combine the
effects of the magnetic hyperfine field (B$_{hf}$) and the electric
field gradient.

\section{Results}
\label{res}

\begin{figure}[h]
\includegraphics*[height=12cm]{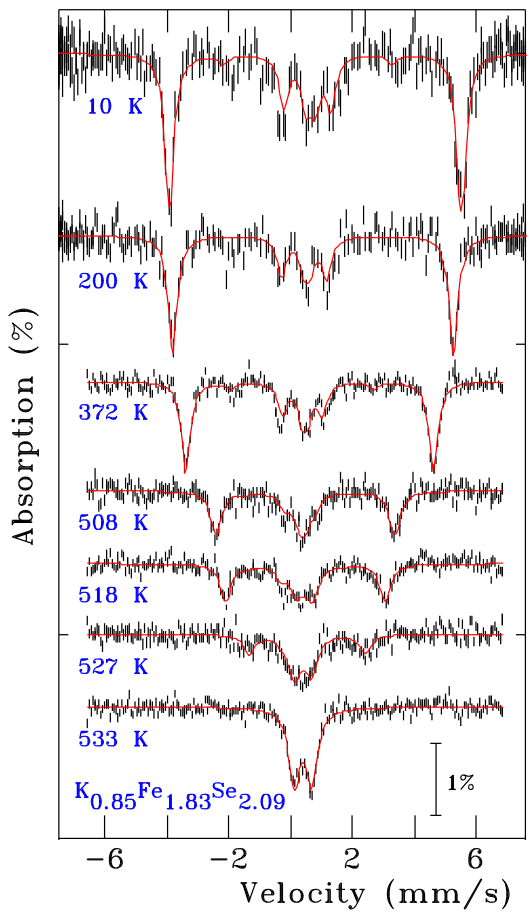}
\caption{\label{fig:spect}
(color online) $^{57}$Fe M\"ossbauer spectra of $\rm K_{0.85}Fe_{1.83}Se_{2.09}$
showing the evolution of the magnetic ordering on heating
from 10~K (well below T$_{sc} \sim 30$~K) to 533~K where the
material is paramagnetic. The absence of the $\Delta m_I$ transitions
in the ordered state indicates that the moments are parallel
to the crystal c-axis (see text), while the growth of a central
paramagnetic component above 500~K is characteristic of a first order
magnetic transition. Solid lines are fits as described in the text.}
\end{figure}

Several conclusions can be reached simply by inspection of the spectrum
taken at 10~K (Fig.~\ref{fig:spect}). The spectrum is dominated by a
single, well-split, magnetic component. This confirms that
$\rm K_{0.85}Fe_{1.83}Se_{2.09}$ is indeed magnetically ordered in the
superconducting state (recall T$_{sc} \sim 30$~K for this sample).
The single magnetic component allows us to rule out any magnetic
structures involving multiple iron sub-sites with moments that differ
by more than a few percent. As we will show below, the large hyperfine
field (B$_{hf} \sim $29~T) is inconsistent with a small iron moment and so
places further limits on possible magnetic structures. Finally, two
of the lines normally present in a magnetically split $^{57}$Fe
M\"ossbauer spectra, are clearly absent from the 10~K pattern.

A magnetic field at the $^{57}$Fe nucleus, either externally applied
or transferred from an ordered moment on the iron atom, lifts the
degeneracy of the nuclear states and, in combination with the
selection rules for the $\frac{3}{2} \rightarrow \frac{1}{2}$
transition, leads to a six-line pattern with intensities of
3:R:1:1:R:3 (counting from left to right in Fig.~\ref{fig:spect}).
For a powder sample, R=2, however if there is a unique angle, $\theta$,
between the magnetic field and the direction of the $\gamma -$beam
used to record the spectrum, then the intensity, R, of the $\Delta m_I = 0$
transitions is given by:
\[
R = \frac{4 \sin ^2 \theta}{1 + \cos ^2 \theta}
\]
R=0 implies that $\theta$ is also zero so that the magnetic field,
and by extension, the moments that lead to it, is parallel to the
$\gamma -$beam. Since the sample consists of an ab-plane mosaic of
single crystals, setting $\theta = 0$ means that the magnetic
ordering direction is parallel to the c-axis, ruling out any
magnetic structures that involve planar ordering of the iron moments.
We note that R is a relatively soft function of $\theta$ near zero, and a free
fit to the intensity of the $\Delta m_I = 0$ transitions is consistent
with an angle of up to 20$^{\circ}$, but this does not lead to a
significant improvement in $\chi ^2$ for the fit, nor would such an
angle be consistent with a planar ordering of the iron moments.
Indeed, if the ordering were planar, then R would be 4, and the
$\Delta m_I = 0$ transitions would provide the strongest features in
the spectrum.

Estimating the iron moment from the observed hyperfine field
requires some care as the scaling is imperfect at best\cite{dubiel}.
However, some data exist on binary iron--chalcogenides that can be
used as a guide (Table~\ref{tab:conv}). If we use the factor of
6.2~T/$\mu_B$ for $\rm Fe_7Se_8$ with our measured B$_{hf}$ of
29.4~T we obtain a rather large estimate of 4.7~$\mu_B$/Fe for
the iron moment in this system. This is significantly larger
than the 3.31~$\mu_B$/Fe reported on the basis of neutron
diffraction\cite{Weibao0830}, however it does suggest that the
iron moment is indeed substantial as even the larger conversion
factor for the sulphide yields 3.5~$\mu_B$/Fe. If we assume that
B$_{hf}$ is at least proportional to the iron moment, then we can
use the observed change in B$_{hf}$ between 10~K and 293~K to
scale the 3.31~$\mu_B$/Fe observed at 11~K\cite{Weibao0830} to
get an estimate of 3.0~$\mu_B$/Fe for the moment at room temperature
for comparison with the much smaller value of 2.55~$\mu_B$/Fe
reported by Pomjakushin {\it et al.}\cite{Pomj3380}. However, the strong temperature
dependence of magnetic signal noted by Bao {\it et al.}\cite{Weibao0830}
suggests a very rapid decline in ordered moment to about 2.8~$\mu_B$/Fe
by room temperature. It is possible that much of the variation
may be intrinsic to the material and its variable stoichiometry,
so that combined measurements on a well
characterised sample will be needed to settle this.

\begin{table}[h]
\caption{Average hyperfine fields
(B$_{hf}$) derived from $^{57}$Fe M\"ossbauer spectroscopy
and average iron moments derived from neutron diffraction
for approximately equi-atomic iron--chalcogenide compounds
with estimated field--moment conversion factors. The Fe--Te
system exhibits significant variability and measurements
have yet to be made on common samples making the conversion
factor unreliable. There is however a clear trend to lower
values in the sequence S$\rightarrow$Se$\rightarrow$Te.
\label{tab:conv}}
\begin{ruledtabular}
\begin{tabular}{lccc}
Compound & Average & Average & Conversion \\
& B$_{hf}$ & moment & Factor \\
& (T) & $\mu_B$/Fe & T/$\mu_B$ \\
\hline
\\
Sulphides \\
$\rm Fe_7S_8$ & 26.8\cite{Kobayashi515} & 3.16\cite{powell014415} & 8.5 \\
\\
\hline
\\
Selenides \\
$\rm Fe_7Se_8$ & 24.1\cite{Ok73} & 3.86 \cite{Andresen64} & 6.2 \\
\\
\hline
\\
Tellurides \\
$\rm Fe_{1.125}Te$ & --- & 2.07\cite{Fruchart169} & \\
$\rm Fe_{1+x}Te$ & --- & 1.96--2.03\cite{Bao247001} & \\
$\rm 0.076 \leq x \leq 0.141$ & & & \\
$\rm Fe_{1.068}Te$ & --- & 2.25\cite{Li054503} & \\
$\rm Fe_{1.05}Te$ &  --- & 2.54\cite{Martinelli094115} & \\
$\rm Fe_{1.11}Te$ & 11\cite{hermon74} & --- & \\
$\rm Fe_{1.08}Te$ & 10.34\cite{MizuguchiS338} & & 4.3--5.2 \\
\end{tabular}
\end{ruledtabular}
\end{table}

Impurities may provide a possible origin for the variation in measured
moments. M\"ossbauer spectroscopy, while sensitive to the presence of
impurity phases, does not rely on normalisation to the total sample in
order to determine moments, they come rather from the observed line splitting,
and not the intensity. Neutron diffraction, by contrast, while providing far
more information on the magnetic ordering, ultimately relies on peak
intensities, normalised to the total nuclear scattering, to determine
the magnetic moments. It is clear from the 10~K spectrum shown in
Fig.~\ref{fig:spect} that there is a central paramagnetic component
present that involves about 12$\pm$2\% of the iron in the sample. Such high
apparent impurity levels in single crystal samples with no impurities
detected by powder x-ray diffraction\cite{Hu1931}, deserves further
attention. If the paramagnetic component is not an ``impurity'' then
it must either be intrinsic to the structure or a property of the material.

At the temperatures of interest here, $\rm K_{0.85}Fe_{1.83}Se_{2.09}$
adopts a vacancy-ordered $I4/m$ modification of the parent $\rm ThCr_2Si_2-$type
$I4/mmm$ structure with iron essentially filling a 16$i$ site and leaving
ordered vacancies on the (almost) empty 4$d$ site \cite{Pomj1919, Zavalij4882}.
Occupations of $\sim$8\% for the Fe-4$d$ site have been reported \cite{Zavalij4882}.
If we assume full occupation of the Fe-16$i$ site in our sample, this leaves
12.5\% of the iron in the 4$d$ site, while scaling the composition to two selenium
atoms per formula unit gives a lower estimate of 9\% for the iron in the
4$d$ site. Partial occupation of the Fe-16$i$ site would leave more iron
to be accommodated in the 4$d$ site. As we see no evidence for a second
magnetic component that could 
be associated with iron in the 4$d$ site, it is possible that the iron in
these more isolated sites does not order, in which case our estimate of
9--12.5\% in the 4$d$ site is fully consistent with the 12$\pm$2\%
paramagnetic component observed in the M\"ossbauer spectrum.

\begin{figure}[h]
\includegraphics*[height=6cm]{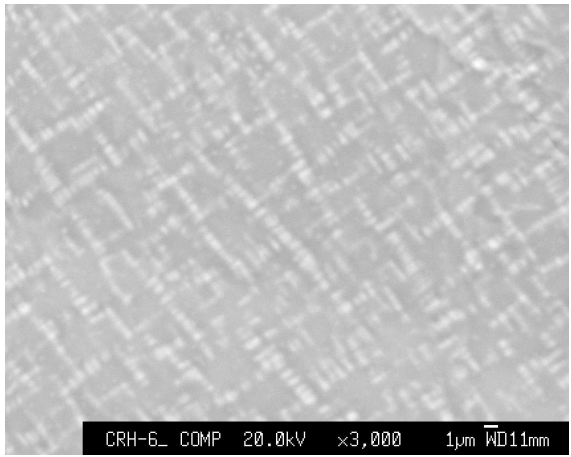}
\caption{\label{fig:bse}
Backscattered electron analysis (BSE) image of a cleaved crystal surface
of $\rm K_{0.85}Fe_{1.83}Se_{2.09}$ taken at an accelerating voltage of 20~kV.
The lighter regions have lower potassium concentrations than the darker
background area.
}
\end{figure}

Another possible origin of the 12$\pm$2\% non-magnetic Fe component in the low
temperature (including room temperature) state can be seen in the
backscattered electron analysis (BSE) image shown in Fig.~\ref{fig:bse}.  This image
reveals that there is, at the micron scale, a modulation in the surface
composition that can be correlated, through a preliminary line-scan analysis of
the WDS data, with reductions of K content in the lighter regions.  Given the
length scale associated with these regions, combined with the probing volume
of the WDS analysis, it is not clear whether this spatial variation is only
associated with the surface or is representative of the bulk behaviour of the
sample.  It should be noted, though, that such patterns appear in samples
grown by furnace cooling as well as samples decanted from a liquid melt\cite{Hu1931}.
Regardless of its origin, the non-magnetic Fe signal
represents a minority of the Fe sites and does not substantially change with
temperature until the first order phase transition is reached.

\begin{figure}[h]
\includegraphics*[height=5cm]{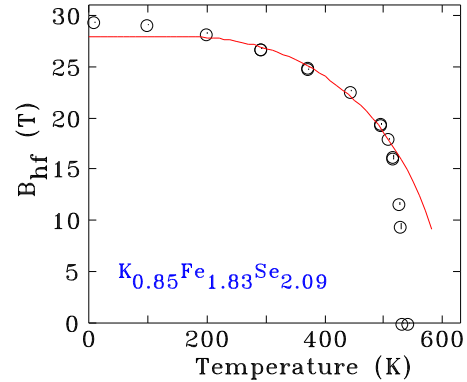}
\caption{\label{fig:bhf}
(color online) Temperature dependence of the magnetic hyperfine field (B$_{hf}$)
in $\rm K_{0.85}Fe_{1.83}Se_{2.09}$. The solid line is a fit to
a J=$\frac{1}{2}$ Brillouin function between 200~K and 500~K
that yields an expected transition of 600$\pm$30~K, well above
the observed value of 532$\pm$2~K. Fitted errors on B$_{hf}$ are
less than 0.1~T, much smaller than the plotting symbols.
The rapid collapse above 500~K
is accompanied by the growth of a paramagnetic component (see
Fig.~\ref{fig:area}).
}
\end{figure}

Raising the temperature leads to the expected decline in 
B$_{hf}$, however it is clear from Fig.~\ref{fig:spect}
that magnetic order persists up to 530~K, confirming that
$\rm K_{0.85}Fe_{1.83}Se_{2.09}$ has a remarkably high ordering
temperature. The temperature dependence of B$_{hf}$ shown in
Fig.~\ref{fig:bhf} yields an ordering temperature of
T$_N= \, 532 \, \pm \, 2$~K. However this is not the result
of the fit to a J=$\frac{1}{2}$ Brillouin function shown in
Fig.~\ref{fig:bhf} as this predicts a transition temperature
of 600$\pm$30~K and the observed behaviour departs from
this curve above 500~K. The two points that bracket the transition
are at 530~K, where a clear magnetic signal is seen, and at 533~K
where the sample is no longer magnetic, setting the transition
at $532\pm2$~K.

A neutron diffraction study of $\rm K_{0.8}Fe_{1.6}Se_{2}$
found two regions in which the temperature dependence of the magnetic
parameter was unusual \cite{Weibao0830}. From 50~K to 450~K
they found a linear dependence of the (101) magnetic peak
intensity, suggesting that $\mu_{Fe}^2$ is a linear function
of temperature. The clear curvature of B$_{hf}$(T) in this
region, shown in Fig.~\ref{fig:bhf}, is not consistent with
this form, as squaring our observed B$_{hf}$(T) to get
something that would scale with the scattering intensity
in a neutron diffraction pattern leads to {\it increased}
curvature rather than linear behaviour.

\begin{figure}[h]
\includegraphics*[height=6cm]{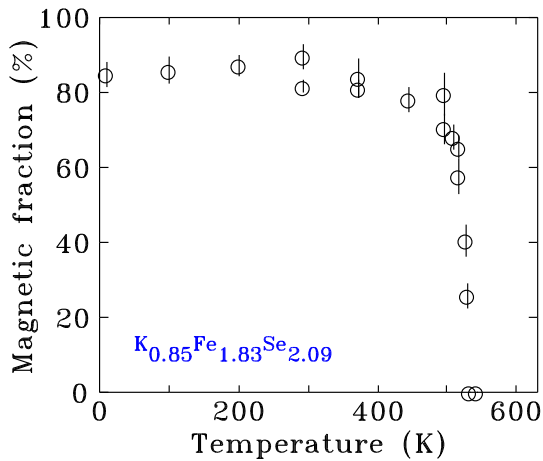}
\caption{\label{fig:area}
(color online) Temperature dependence of the magnetic fraction
in $\rm K_{0.85}Fe_{1.83}Se_{2.09}$. The rapid collapse
above 500~K indicates that the magnetic transition has
first order character and may be associated with a structural
transition.
}
\end{figure}

Above 500~K, Wei Bao {\it et al.} reported a very rapid
decrease in the (101) intensity \cite{Weibao0830} leading
to an ordering temperature of $\sim$560~K. While our
sample composition is slightly different and our ordering
temperature slightly lower, we see the {\it same} abrupt
loss of magnetic order in Fig.~\ref{fig:bhf}. Inspection of
the spectra above 500~K shown in Fig.~\ref{fig:spect} reveals
that the intensity of the magnetic peaks decreases visibly
as their splitting falls. The ability to uniquely separate
the amount of a magnetic phase (seen through line intensities)
from the magnitude of the magnetic order (seen independently
through line splittings) is an important strength of
M\"ossbauer spectroscopy. Tracking the fraction of the iron
that is present as a magnetically ordered form (Fig.~\ref{fig:area})
confirms that the magnetic phase is disappearing even faster
than the splitting that marks the order. This strongly
suggests that the magnetic phase is transforming before it
reaches its true ordering temperature (which we estimate to
be about 600~K) and that the observed transition is being
driven by a first order structural event. This view is
supported by the neutron diffraction work of Wei Bao {\it et al.}
\cite{Weibao0830} where they also tracked the intensity of
the (110) structural peak that is associated with the $I4/m$
vacancy-ordered structure of $\rm K_{0.8}Fe_{1.6}Se_{2}$
below 580~K. This peak starts to lose intensity at the same
temperature at which the (101) magnetic peak starts its sudden
decline. As we see both a weakening of the magnetic order
and a reduction in the magnetic fraction above 500~K, it is
possible that the break-up of the vacancy-ordered magnetic
form reduces the magnetic connectivity of the ordered phase
until it forms a non-percolating network of finite clusters.
The magnetic order is then lost at a temperature below both
its intrinsic ordering temperature, and the temperature at
which the vacancy-ordered $I4/m$ structure fully transforms
to the high-temperature $I4/mmm$ form.

\section{Conclusions}
\label{conc}

Our $^{57}$Fe M\"ossbauer spectroscopy study of single crystals
of $\rm K_{0.85}Fe_{1.83}Se_{2.09}$ confirms the presence of magnetic
order from well below T$_{sc} \sim 30$~K to T$_N= \, 532 \, \pm \, 2$~K.
The large magnetic splitting of 29.4$\pm$0.1~T at 10~K indicates that
the iron moments are large, consistent with values of 3.31~$\mu_B$/Fe
observed by neutron diffraction at 11~K\cite{Weibao0830}, while the
line intensities indicate that the ordering is parallel to the c-axis.
An apparent paramagnetic impurity phase is attributed to iron atoms
in the 4$d$ site. Analysis of the spectra taken in the vicinity of
T$_N$ shows that the magnetic fraction decreases rapidly above 500~K
and that the loss of order is driven by a first order structural
transition as the material transforms into the high-temperature $I4/mmm$ form.

\section*{Acknowledgements}

Financial support for various stages of this work was provided by
the Natural Sciences and Engineering Research Council of Canada
and Fonds Qu\'eb\'ecois de la Recherche sur la Nature et les Technologies.
JMC acknowledges support from the Canada Research Chairs programme.
R.H. and P.C.C. are supported by AFOSR--MURI grant \#FA9550-09-1-0603.
W.E.S., S.L.B. and P.C.C. are also supported by the U.S. Department of Energy,
Office of Basic Energy Science, Division of Materials Sciences and
Engineering. Ames Laboratory is operated for the U.S. Department
of Energy by Iowa State University under Contract No. DE-AC02-07CH11358.

\end{document}